\documentclass[%
 reprint, nofootinbib,superscriptaddress,twocolumn 
 amsmath,amssymb,
pra,
]{revtex4-2}

\usepackage{tabularx}
\usepackage{graphicx}
\usepackage{xcolor}
\usepackage{dcolumn}
\usepackage{bm}

\newcommand{\ket}[1]{\left|#1\right\rangle}
\newcommand{\bra}[1]{\left\langle #1\right|}

\usepackage{amsmath}
\usepackage{wrapfig}
\usepackage{graphicx}
\usepackage{float}
\usepackage{subfigure}
\usepackage{amssymb}
\usepackage{bbold}
\usepackage{color}
\usepackage{ulem}
\usepackage{comment}

\begin{document}

\preprint{APS/123-QED}

\title{Quantum walk processes in quantum devices}

\author{Anandu~Kalleri~Madhu}
\affiliation{Department of Physics, National Tsing Hua University, Hsinchu 30013, Taiwan}

\author{Alexey~A.~Melnikov}
\thanks{Corresponding author, e-mail: melnikov@phystech.edu}
\affiliation{Valiev Institute of Physics and Technology, Russian Academy of Sciences, 117218 Moscow, Russia}

\author{Leonid~E.~Fedichkin}
\affiliation{Valiev Institute of Physics and Technology, Russian Academy of Sciences, 117218 Moscow, Russia}

\author{Alexander~P.~Alodjants}
\affiliation{ITMO University, 197101 St. Petersburg, Russia}
\affiliation{Quantum Light Engineering Laboratory, Institute of Natural and Exact Sciences,
South Ural State University (SUSU), 76, Lenin Av., Chelyabinsk, Russia}

\author{Ray-Kuang~Lee}
\affiliation{Department of Physics, National Tsing Hua University, Hsinchu 30013, Taiwan}
\affiliation{Institute of Photonics Technologies, National Tsing Hua University,  Hsinchu 30013, Taiwan}
\affiliation{Physics Division, National Center for Theoretical Sciences, Taipei 10617, Taiwan}
\affiliation{Center for Quantum Technology, Hsinchu 30013, Taiwan}


\begin{abstract}
Simulation and programming of current quantum computers as Noisy Intermediate-Scale Quantum (NISQ) devices represent a hot topic at the border of current physical and information sciences. The quantum walk process represents a basic subroutine in many quantum algorithms and plays an important role in studying physical phenomena. Simulating quantum walk processes is computationally challenging for classical processors. With an increasing improvement in  qubits fidelity and qubits number in a single register, there is a potential to improve quantum walks simulations substantially. However, efficient ways to simulate quantum walks in qubit registers still have to be explored. Here, we explore the relationship between quantum walk on graphs and quantum circuits. Firstly, we discuss ways to obtain graphs provided quantum circuit. We then explore techniques to represent quantum walk on a graph as a quantum circuit. Specifically, we study hypercube graphs and arbitrary graphs. Our approach to studying the relationship between graphs and quantum circuits paves way for the efficient implementation of quantum walks algorithms on quantum computers.   
\end{abstract}

\maketitle

\section{Introduction}

Random walks on graphs naturally appear in different physical processes~\cite{kac1947random,bartumeus2005animal,brockmann2006scaling,codling2008random} and computational subroutines~\cite{rajeev1995randomized,landau2001efficient,sottinen2001fractional,sabelfeld2013random,gkantsidis2006random}. In order to study these physical processes and implement algorithms, there is a need for efficient ways to simulate random walks~\cite{gillespie1978monte,cohen2016faster}. Quantum walks, quantum analogues of (classical) random walks~\cite{aharonov1993qw,kempe2003qw,konno2008quantum,venegas2012quantum,PhysRevA.58.915,manouchehri2013qw,fedichkin2020analysis}, naturally appears in physical processes when studied at a quantum level. Quantum interference, which is at the heart of quantum walks~\cite{PhysRevA.73.032341,solenov2006continuous,fedichkin2006mixing,melnikov2016quantum,su2019experimental,Cui2020quantum}, potentially enables to accelerate energy transfer in Fenna-Matthews-Olson complexes~\cite{engel2007evidence,mohseni2008environment,Harel} and quantum photonic circuits~\cite{fulvio2018photonic, alodjants2022photonic}. Understanding quantum walk advantage in particle transfer requires efficient simulation techniques and graph classification algorithms~\cite{melnikov2019predicting,melnikov2020machinetransfer,kryukov2022supervised}. Simulating quantum walks is computationally a \#P-hard problem, which makes classical simulators inefficient for the task~\cite{qiang2016efficient,acasiete2020implementation}. In this regard, efficiently simulating quantum walks in quantum devices, especially in quantum computers available now represents an important task for understanding quantum transport advantages~\cite{harris2017quantum,maier2019environment}, and for quantum algorithms implementation~\cite{chen2019hybrid} as quantum walks represent a universal model of quantum computation~\cite{Childs1,Childs2}. 

However, existing quantum computers are NISQ devices, and possess limited capabilities due to a small number of qubits~\cite{preskill2018quantum}. In this sense, the simulation of quantum walks on large-scale graphs with such devices represents an important and non-trivial task.

In this work, we study the practical possibilities to run Continuous-Time Quantum Walk (CTQW) processes in NISQ devices. The paper is mainly structured into two main parts as shown in Fig.1. In the first section, we discuss about how to obtain graphs given a quantum circuit. We talk about two different kinds of graphs: i) hypercubes and ii) arbitrary graphs. In the second section, we discuss how to obtain a quantum circuit given a graph. We talk about how to simulate the CTQW on a graph using a quantum circuit. In particular, we discuss simulating hypercube graphs and arbitrary graphs. As an example of simulating an arbitrary graph, we simulate a 4 node ``paw" graph on a quantum computer. We also discuss the technique of Hamiltonian simulation for simulating arbitrary unitary state evolution in a quantum computer.

\begin{figure}[ht!]
	\centering
	\includegraphics[width=1\linewidth]{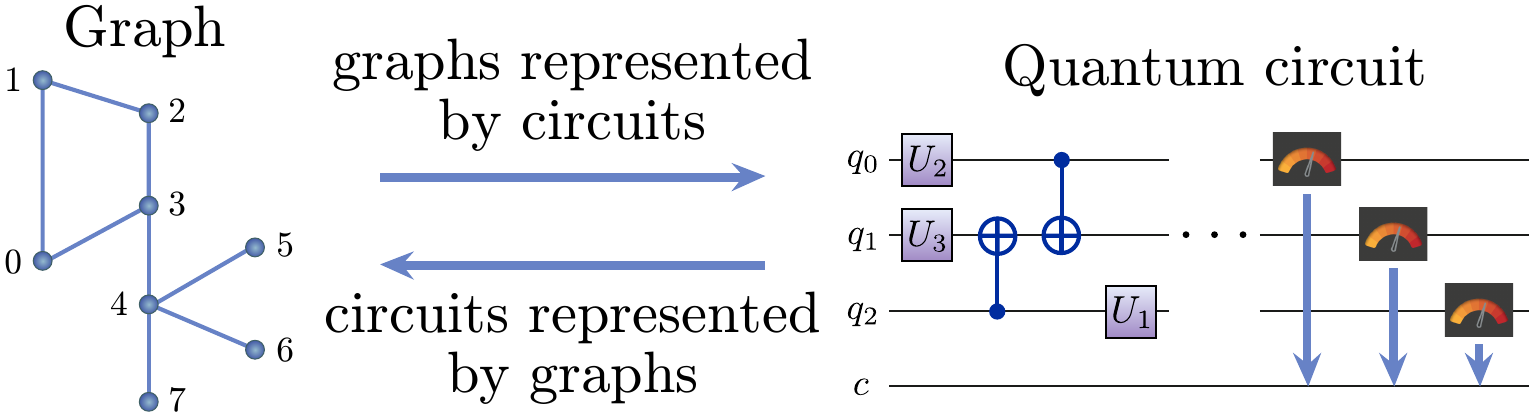}
	\caption{A schematic relation between graphs and quantum circuits.}
	\label{fig1}
\end{figure}

\section{Simulating quantum walks on graphs}

Single particle CTQW on graphs are studied in this paper. A quantum particle can be located in one of the $d$ positions on a graph with $d$ vertices, or in a superposition of these positions. A quantum state of this particle can be thought of as a state of a $d$-level system as shown in Eq.~(\ref{quantum_state_dlevel}):

\begin{equation}
   \ket{\psi_d(t)} = \sum_{i=0}^{d-1} \alpha_i(t)\ket{i} 
   \label{quantum_state_dlevel}
\end{equation}
where $\lvert\alpha_i(t)\rvert^2$ being the detection probability in vertex $i$ at a time $t$ and verifying in Eq.~(\ref{prob_verification}):
\begin{equation}
    \sum_{i=0}^{d-1} \lvert\alpha_i(t)\rvert^2 = 1
    \label{prob_verification}
\end{equation}
The evolution of this quantum state is governed by the Hamiltonian with nearest-neighbour hopping terms given by Eq.~(\ref{Hamiltonian_nearestneighbour}):
\begin{equation}
    \mathcal{H}_A = \hbar\Omega\sum_{i,j=0}^{d-1} A_{ij}\ket{i}\bra{j} = \hbar\Omega A, 
    \label{Hamiltonian_nearestneighbour}
\end{equation}
where $A$ is an adjacency matrix of a graph on which the quantum walk is performed, $A_{ij}$ are the elements of this matrix and $\Omega$ is the hopping frequency. 

The unitary quantum state evolution is hence a solution to the Schr\"{o}dinger equation, which is given by
\begin{equation}
    \ket{\psi_d(t)} = \mathrm{e}^{-i\Omega t A}\ket{\psi_d(0)}.
    \label{general_evolution}
\end{equation}
Note that $A$ is not necessarily symmetric -- the weights $A_{ij}$ are complex parameters and can, in general, lead to a chiral quantum walk on a weighted graph~\cite{zimboras2013quantum,lu2016chiral}.

Exponentiation of the $\Omega tA$ matrix is computationally challenging for large $d$. In the case of using a quantum computer, however, it is known that some unitary matrices can be efficiently implemented in time logarithmic in $d$. However, as the form of a Hermitian matrix $A$ is arbitrary, we cannot efficiently implement any quantum walk on a quantum computer. In the general case, the matrix $A$ is described by $(d^2-d)/2$ independent complex-valued variables, which means in the worst case, one would need to apply $O(d^2)$ operations making exponential speedup impossible. In this paper, we explore the set of matrices $A$, for which efficient implementation is possible, and demonstrate this implementation on the IBM Q quantum device.

\section{Obtaining graphs given quantum circuits}
\label{GraphFromCircuits}

\begin{figure*}[ht!]
	\centering
	\includegraphics[width=1\linewidth]{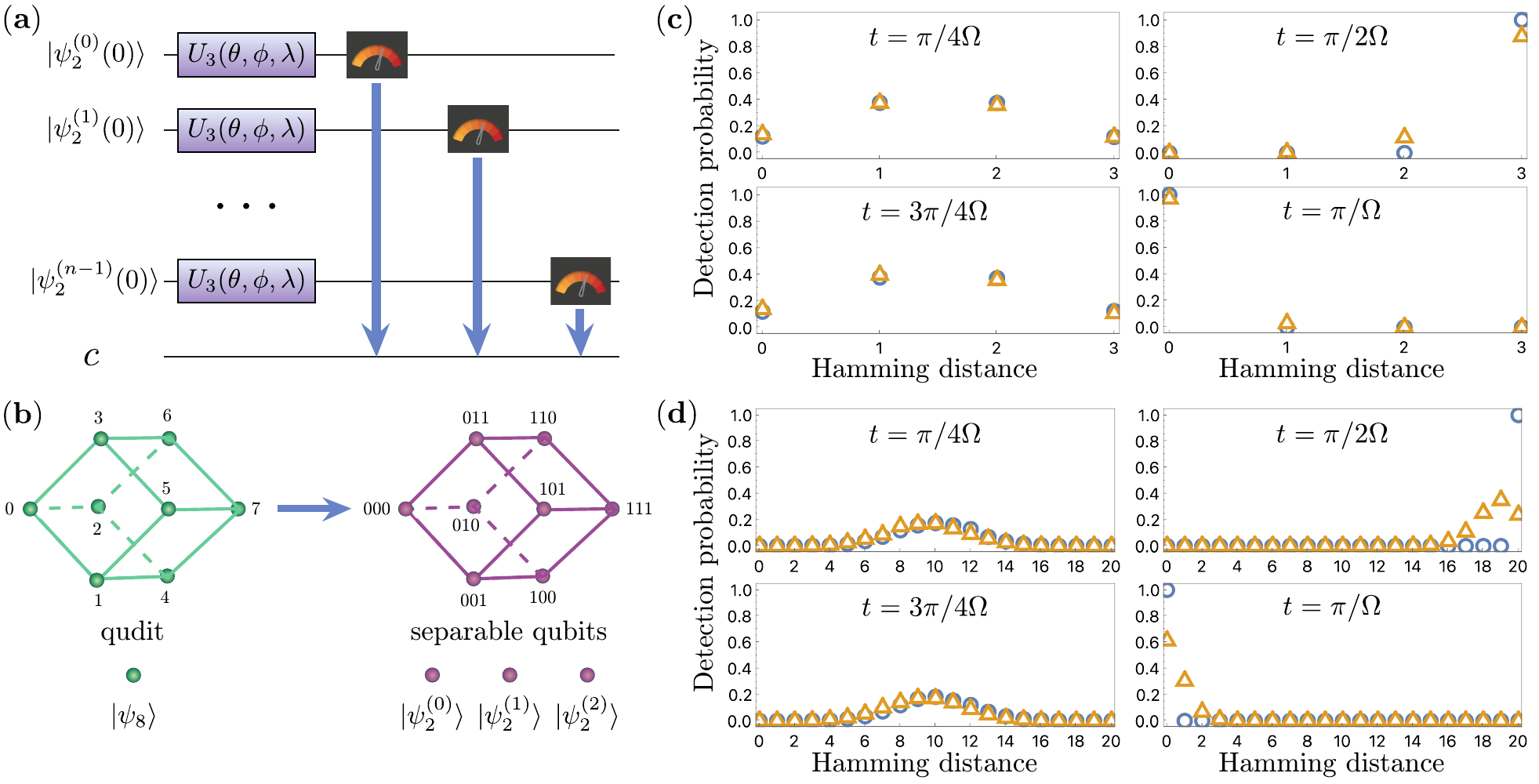}
	\caption{(a) Quantum circuit for a CTQW implemented on $n$ separable qubits. (b) Two different ways to encode a state of a walking particle on a hypercube graph: one qudit (left), and $n$ separable qubits (right). Quantum and classical simulation of CTQW on(c) a $3$-dimensional hypercube and (d) a $20$-dimensional hypercube. Each plot shows a probability distribution of particle position on a graph at different times, corresponding to quantum circuit parameters $\phi=-\pi/2, \lambda=\pi/2$ and $\theta = 2\Omega t=\pi/2$, $\pi$, $3\pi/2$, $2\pi$. The blue circles and orange triangles correspond to execution on the classical simulator and the quantum IBM Q device, respectively. Hamming distance indicates the distance at which the particle propagated starting from the initial state labeled as ``$00\dots 0$''.}
	\label{fig2}
\end{figure*}

In this section, we discuss how to obtain graphs given a quantum circuit. A quantum circuit corresponds to a unitary state evolution of a Hamiltonian. This Hamiltonian represents the CTQW of a particle on a graph. The adjacency matrix of this graph is obtained from the underlying unitary represented by the quantum circuit. Therefore, every quantum circuit corresponds to a graph on which CTQW are implemented. In this section, we briefly discuss obtaining two families of graphs: i) hypercubes; ii) arbitrary graphs. 

\subsection{Quantum walks on hypercubes}

Hypercube graphs represent a starting point towards studying graphs with a large number of vertices connected non-trivially. To simulate quantum walks on hypercubes, one needs to establish a mapping. Firstly, an encoding of the particle's position in the space of qubits needs to be defined. Secondly, a sequence of quantum gates must be specified, which corresponds to a unitary defined by an adjacency matrix.
Here we consider two possibilities of mappings, although there could be more, which correspond to an exponential reduction of the number of qubits compared to the graph vertices. The first possibility is based on using separable qubits, whereas the second possibility requires entangled qubits. In both cases the state $\ket{\psi_d(t)}$ is simulated on a classical simulator and IBM Q quantum devices.

The first mapping is demonstrated in Fig.~\ref{fig2}(b). Instead of describing a quantum particle's position as a state of a single $d$-level quantum system, we can think of the position as a state of an $n=\log_2 d$ qubits system. These two different ways of encoding a quantum walker state are shown in Fig.~\ref{fig2}(b) for cube graphs. Both cube graphs in Fig.~\ref{fig2}(b) have the same adjacency matrix $A^\mathrm{hc}$. The two graphs' difference is only in the vertex labeling: decimal (left cube, from $0$ to $7$) and binary (right cube, from $000$ to $111$) encoding choices. The binary encoding choice replaces a $d$-level quantum system, qudit, with multiple $2$-level systems, qubits. 

In addition to the quantum walkers state mapping, we also specify the mapping for unitary operations defined by the graph edges. The binary labelling is shown in Fig.~\ref{fig2}(b) on the right helps us to see that two vertices are connected by a bit-flip operation. Moreover, each $i$-th bit-flip corresponds to a walk on the $i$-th axis. Therefore, a walk on a hypercube is a sequence of independent bit-flip operations. In other words, CTQW on an arbitrary $n$-dimensional hypercube can be decomposed into CTQW on $n$ independent line graphs. The same can be observed by decomposing the $A^\mathrm{hc}$ matrix as in Eq.~(\ref{Adj_decomposition}):
\begin{equation}
    A^\mathrm{hc} = \sum_{i=0}^{n-1} I_2^{\otimes i}\otimes A_i^\mathrm{line}\otimes I_2^{\otimes n-1-i}.
    \label{Adj_decomposition}
\end{equation}
Moreover, as a direct consequence of the above identity, the unitary matrix that is a function of $A$ is equal to
\begin{equation}
    U(A^\mathrm{hc}) = \mathrm{e}^{-i\Omega t A^\mathrm{hc}} = \left(\mathrm{e}^{-i\Omega t A^\mathrm{line}}\right)^{\otimes n},
    \label{unitary_hypercube}
\end{equation}
where $A^\mathrm{line}=\ket{0}\bra{1}+\ket{1}\bra{0}$ represents an adjacency matrix of a two-vertex line graph.

We next implement the CTQW on $n$-dimensional hypercubes in qubit registers of IBM Q. Because of the unitary operations simplified form, obtained in Eq.~(\ref{unitary_hypercube}), we implement $n$ identical CTQW on a line. Each quantum walk on the line is simulated by a single-qubit unitary that evolves the qubit state $\mathinner{\lvert \psi_2^{(i)}\rangle}$ shown in Fig.~\ref{fig2}(a). For the simulation, we use the $U_3$ gate given by Eq.~(\ref{U3}) in Qiskit to evolve the CTQW during the time $t=\theta/2\Omega$:
\begin{equation}
    U_3(\theta,\phi,\lambda) = \begin{bmatrix}
    \cos{\frac{\theta}{2}} &~~-\mathrm{e}^{i\lambda}\sin{\frac{\theta}{2}}\\\\
    \mathrm{e}^{i\phi}\sin{\frac{\theta}{2}} &~~\mathrm{e}^{i(\lambda+\phi)}\cos{\frac{\theta}{2}}
    \end{bmatrix}.
    \label{U3}
\end{equation}
A combined CTQW circuit, which is composed of the $U_3$ gates is depicted in Fig.~\ref{fig2}(a). The initial position of the simulated particle is defined by the initial quantum state $\ket{\psi_{2^n}(0)}=\mathinner{\lvert \psi_2^{(1)}(0)\rangle}\otimes\dots\otimes\mathinner{\lvert \psi_2^{(n)}(0)\rangle}$. Because of the symmetry of hypercube graphs, without the loss of generality, the initial vertex will always be ``$00\dots0$'' and $\ket{\psi_{2^n}(0)}=\ket{00\dots 0}$. Simulating the evolution of $\ket{\psi_{2^n}(t)}$ with the specified initial condition, we observe the detection probabilities $p_k = \lvert\mathinner{\langle k\lvert \psi_{2^n}(t)\rangle} \rvert^2$ by repeating the circuit execution on the IBM Q quantum device.

As a result of the simulations, we obtained the particle's probability to be in different vertices of the hypercubes in different time steps. The results for graphs with $n = 3$ and $n = 20$ are given in Fig.~\ref{fig2}(c) and Fig.~\ref{fig2}(d), respectively. The results demonstrate that starting from the ``$00...0$'' initial node, the quantum particle moves along the graph reaching the furthest node ``$11...1$'' with almost unit probability at time $t=\pi/2\Omega$. The same holds for both $n=3$ and $n=20$. Indeed, for any dimension $n$, the time for a perfect particle transfer is constant $t=\pi/2\Omega$. This is expected since we evolved the walk through separable qubits independently. 
In the results shown in Fig.~\ref{fig2}(c)-(d), classical simulated results are shown as blue circles and quantum simulated results are shown as orange triangles. We observe that for times $t = \pi/4\Omega$ and $t = 3\pi/4\Omega$ the quantum simulated probability distribution matches the classical simulated probability distribution with an error in probability below $0.05$ for all data points. In cases of $t = \pi/2\Omega$ and $t = \pi/\Omega$, however, the errors in quantum simulation go up to $0.15$ for $n=3$, and up to $0.8$ for $n=20$. In addition, the errors are counter-intuitively lower for $t = \pi/\Omega$ compared to $t = \pi/2\Omega$, which is explained by the underlying quantum system symmetry. However, in the case of separable qubits, the quantum simulation does not provide any more advantage than classical simulation since we haven’t used any entangled qubits for the simulation.

\begin{figure*}[ht!]
	\centering
	\includegraphics[width=1\linewidth]{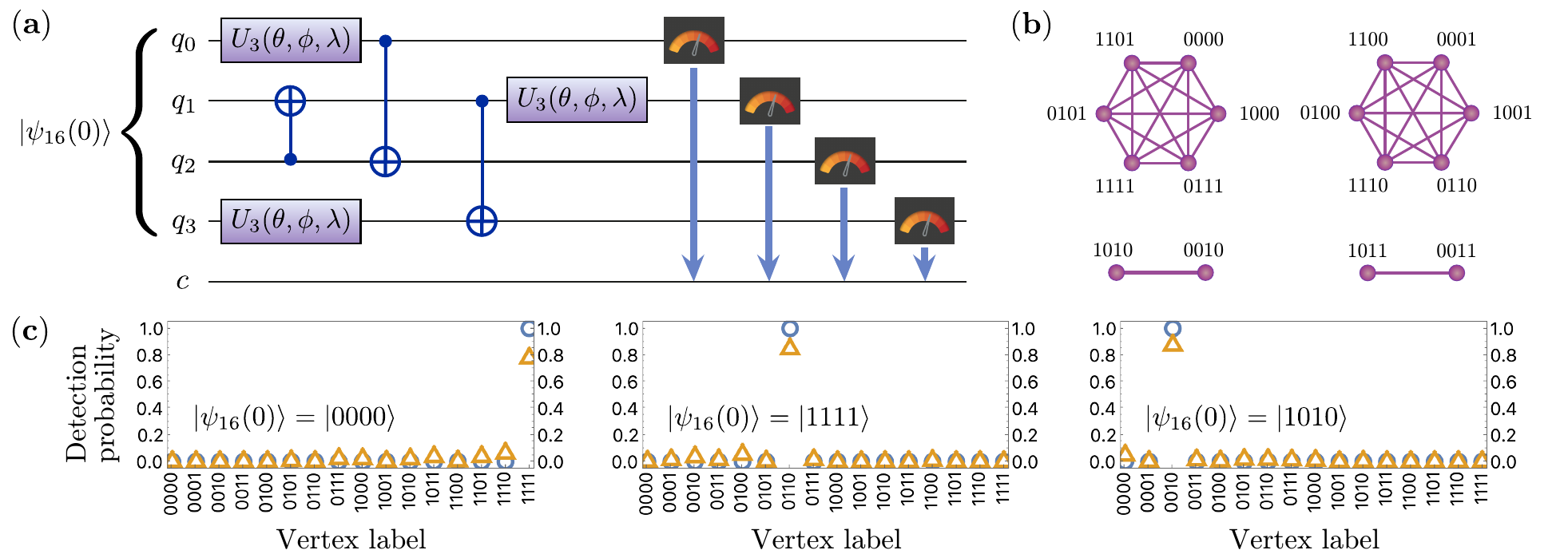}
	\caption{(a) A $4$-qubit quantum circuit consisting of $U_3(\theta = \pi, \phi=-\pi/2, \lambda=\pi/2)$ and CNOT gates with a circuit depth of $5$. (b) A weighted graph generated from a $4$-qubit quantum circuit with $U_3$ and CNOT gates on which we evolved a CTQW starting from the node ``$0000$'' and ending up in ``$1111$'' with probability one. Weights are not shown. (c) Quantum and classical simulation results for a quantum walk on the graph (b). Each plot shows a probability distribution of particles position on a graph for different initial conditions: $\ket{\psi_{16}(0)}=\ket{0000}$, $\ket{1111}$, and $\ket{1010}$. The blue circles and orange triangles correspond to execution on the simulator and the IBM Q quantum register, respectively. Vertex labels correspond to vertices in (b).}
	\label{fig3}
\end{figure*}

\subsection{Quantum walks on arbitrary graphs}

Not all unitary operations are efficiently implementable on a quantum computer, hence not all graphs $A$ with $d$ vertices can be implemented with $O\left(\log d\right)$ qubits in time $O\left(\mathrm{poly}(\log d)\right)$. Nonetheless, as we demonstrated in the previous section, for the hypercube graphs it is possible. By studying CTQW on hypercubes, and providing different implementations of the same process, we conclude that some implementations are more feasible than others. To see what other graphs have a feasible and efficient implementation on a quantum device, we propose the following.

To find out which graphs can be efficiently implemented on a quantum computer, we explore different quantum circuits that are efficiently implementable on the IBM Q quantum computer. Each of these quantum circuits represent some specific unitary operation $U$, which has a corresponding adjacency matrix $A(U)$. Because of the form of evolution $U=\mathrm{e}^{-i\Omega t A}$ in Eq.~(\ref{general_evolution}), $A(U)=i\log(U)/\Omega t$. The logarithm of a unitary matrix is not uniquely defined, however~\cite{loring2014computing}. First, adding a term $2\pi kI$, $k\in \mathcal{Z}$ with $I$ being the identity matrix does not introduce a change in matrix $A$. Second, adding a global phase $c\in\mathcal{C}$ to a unitary $U'=cU$ does not change the particle's quantum walk dynamics. The freedom in phase leads to the possibility to add a term $\varphi I$, $\varphi\in\mathcal{R}$ to any adjacency matrix without affecting the quantum walk simulation. Third, one adds a complete graph with a factor $2\pi k A_\mathrm{complete}/n\Omega t$, $k\in\mathcal{Z}$, without introducing a change in $U$. This is a consequence of $A_\mathrm{complete}/n$ being an idempotent matrix with $n$ -- number of qubits. This condition is only valid if $A_\mathrm{complete}/n$ commutes with U. Finally, combining all three possibilities to modify $A$ together, we obtain
\begin{equation}
    A = \frac{i}{\Omega t}\log(U) + \varphi I + \frac{2\pi k}{n\Omega t} A_\mathrm{complete},
    \label{AfromU}
\end{equation}
with free parameters $k\in\mathcal{Z}$, and $\varphi\in \mathcal{R}$. In addition to the derived Eq.~(\ref{AfromU}), multiplying $A$ by an arbitrary factor $b\in \mathcal{R}$ leads to an effective rescaling of the transition frequency $\Omega'=\Omega/b$, which broadens the set of available $A$ even more.

\begin{figure*}[ht!]    
	\centering
	\includegraphics[width=1\linewidth]{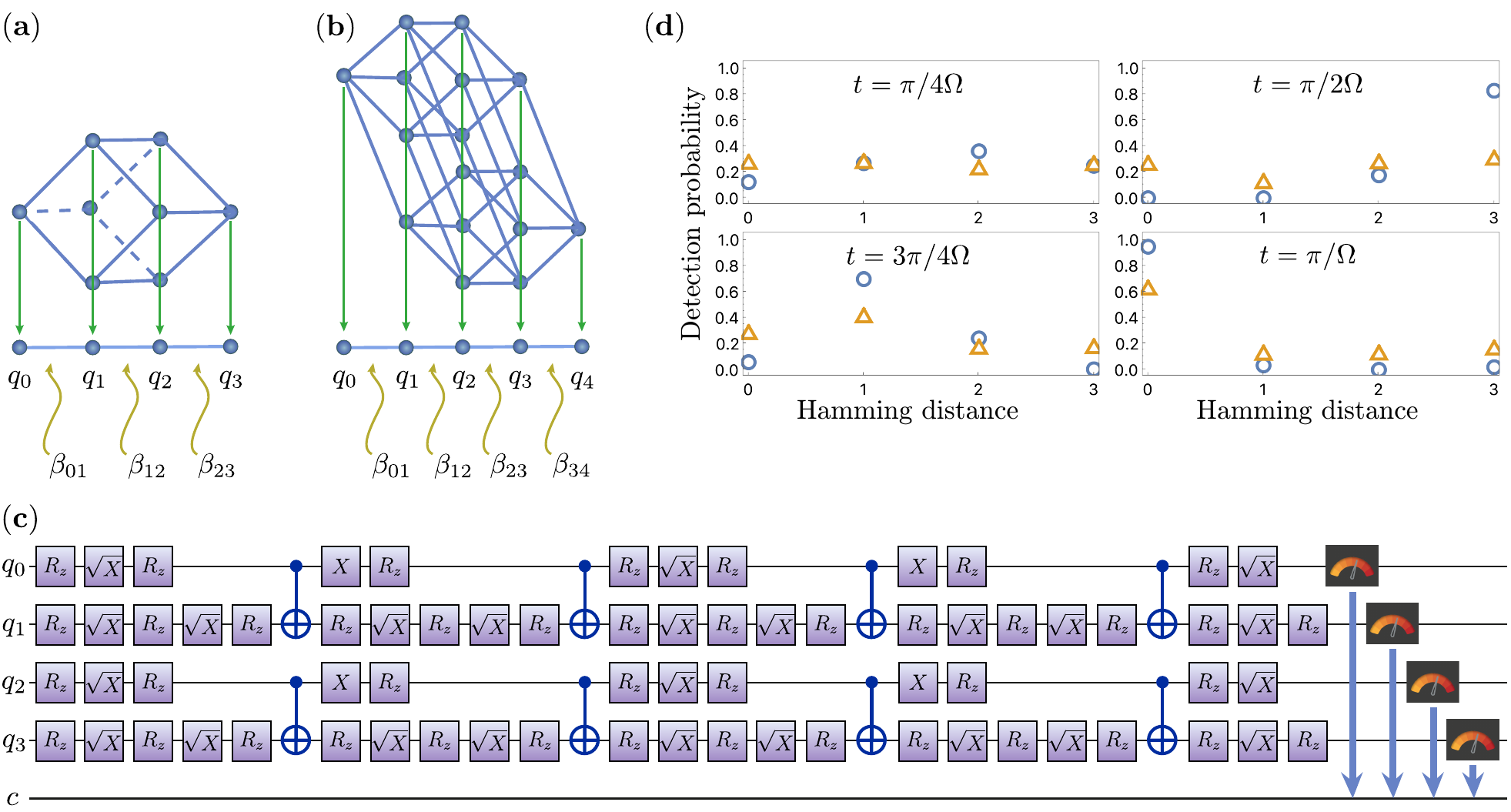}
	\caption{(a)-(b) Mapping an $n$-dimensional hypercube graph to a line graph. Examples for $n=3$ (a) and $n=4$ (b) are shown. Weights $\beta$ are defined in Eq.~(\ref{hypercube_to_line}). (c) Quantum circuit for a CTQW implemented for $t=\pi/2\Omega$ on four entangled qubits. (d) Quantum and classical simulation results for implementing a CTQW on a weighted line with $4$ vertices. Each plot shows a probability distribution of particles position on a graph at different times: $t=\pi/4\Omega$, $\pi/2\Omega$, $3\pi/4\Omega$, and $\pi/\Omega$. The blue circles and orange triangles correspond to execution on the simulator and the IBM Q quantum register, respectively. Hamming distance indicates the distance at which particle propagated starting from the initial state corresponding to an excited state of the $q_0$ qubit.}
	\label{fig4}
\end{figure*}

The derived Eq.~(\ref{AfromU}) helps us to obtain a variety of adjacency matrices of potential interest given the unitary transformation. Our implementation of the unitary transformations, in turn, is adjusted to the quantum device's connectivity. From the quantum device's connectivity limitations, it is possible to simulate which $U(A)$ are easily implementable with $k$-depth quantum circuits. For this, we automated sampling of random circuits that implement perfect transport from the state ``$0\dots 00$'' to the state ``$1\dots 11$'' with a circuit depth up to $5$. We obtained these circuits to find feasible $U$, and from it, feasible graphs $A(U)$.
In the case of $4$ qubits, we obtained a graph shown in Fig.~\ref{fig3}(b), by randomly implementing $U_3$ and CNOT gates. The corresponding quantum circuit is given in Fig.~\ref{fig3}(a). Given that the initial state is ``$0000$'', the circuit's transformation is equivalent to unitary in Eq.~(\ref{unitary_hypercube}). However, for the general initial particle's position, i.e., general $\ket{\psi_{16}(0)}$, the transformation is different from the hypercubes cases. This difference can be observed from the disconnected graph shown in Fig.~\ref{fig3}(b) compared to a fully connected graph of a $4$-dimensional cube in Fig.~\ref{fig4}(b). Quantum and classical simulation results in Fig.~\ref{fig3}(c) demonstrate the chiral nature of the graph. Indeed, starting from $\ket{\psi_{16}(0)}=\ket{0000}$, particle ends in $\ket{1111}$, however starting from $\ket{\psi_{16}(0)}=\ket{1111}$ brings the particle to the ``$0110$'' vertex. The simulations performed for the initial state $\ket{\psi_{16}(0)}=\ket{1010}$ shows that the particle is bounded to the subspace of the ``$1010$'' and ``$0010$'' vertices. Note that the fidelities of all the quantum simulations in Fig.~\ref{fig3} are above $0.75$.

\section{Obtaining quantum circuits given graphs}
\label{CircuitsFromGraphs}

In the section above, we have discussed how to obtain graphs given quantum circuits. In this section, we discuss obtaining quantum circuits corresponding to CTQW of a particle on a graph. The adjacency matrix $A$ of a graph corresponds to the Hamiltonian of the CTQW of a particle on that graph. Therefore, to represent a graph as a quantum circuit, we simulate the unitary state evolution of the Hamiltonian corresponding to the adjacency matrix of the graph. In this section, we discuss how to simulate: i) hypercube graphs; ii) arbitrary graphs on a quantum computer using quantum circuits.

\subsection{Quantum walks on hypercubes}

A different mapping can be used to simulate a CTQW on a hypercube. In this mapping, one exploits the fact that there is an equal probability of detecting a particle in vertices with the same Hamming distance. By taking all the symmetries into account, one obtains a weighted line graph. The mapping procedure is shown in Fig.~\ref{fig4} for $n=3$ (a) and $n=4$ (b). To construct the Hamiltonian that governs the time evolution of the quantum walk in the mapped space, we use $XY$ coupling terms, where:
\begin{equation}
    X = \begin{bmatrix}
    0 &~~1\\\\
    1 &~~0
    \end{bmatrix},
    \label{X_gate}   
\end{equation}
and 
\begin{equation}
    Y = \begin{bmatrix}
    0 &~~-i\\\\
    i &~~0
    \end{bmatrix},
    \label{Y_gate}   
\end{equation}
are the Pauli matrices as represented in Eq.~(\ref{X_gate}) and Eq.~(\ref{Y_gate}) respectively. The nodes of the weighted line graph are implemented as qubits and are expressed as $X$ and $Y$ terms in our Hamiltonian. The weights of the edges of the graph are specified by the strength of the couplings between qubits. In our Hamiltonian, we represent them as coefficients of the $XY$ coupling terms $\beta_{i,i+1}=\sqrt{(1+i)(n-i)}$, where n is the dimension of the hypercube and the indices \textit{i} represent qubits or nodes of the graph. The Hamiltonian that governs the quantum walk in the mapped space can be defined for an arbitrary $n$:
\begin{equation}
    H^\mathrm{hc\rightarrow line} = \frac{1}{2} \sum_{i=0}^{n-1}\beta_{i,i+1}\left(X_i X_{i+1} + Y_i Y_{i+1}\right),
    \label{hypercube_to_line}
\end{equation}
Implementing a quantum walk in this mapping has an advantage of the exponential reduction of the number of qubits, as there are only $(n+1)$ qubits needed for $n$-dimensional hypercube CTQW implementation. This is similar to the mapping with separable qubits considered before, which required $n$ qubits.

For the simulation of CTQW on a hypercube with this mapping, we have to implement the above Hamiltonian in Eq.~(\ref{hypercube_to_line}) on the IBM quantum computer. Here, we simulate a hypercube of dimension $n=3$ mapped to a weighted line graph. The corresponding Hamiltonian for the weighted line graph can be obtained from Eq.~(\ref{hypercube_to_line}) for $n=3$ as shown in Eq.~(\ref{hypercube_n=3_to_line}),

\begin{multline}
H^\mathrm{hc\rightarrow line} = \frac{\sqrt{3}}{2}(X_0 X_1 + Y_0 Y_1) + (X_1 X_2 + Y_1 Y_2) + \\ \frac{\sqrt{3}}{2}(X_2 X_3 + Y_2 Y_3),
\label{hypercube_n=3_to_line}    
\end{multline}

The $16\times16$ adjacency matrix of the line graph corresponds to this Hamiltonian. Since the terms of the Hamiltonian do not commute, we use Trotter decomposition to simulate the Hamiltonian. Trotter decomposition can be used to accurately simulate the Hamiltonian's unitary time evolution by breaking it up into a series of short time-steps as shown in Eq.~(\ref{trotter})

\begin{equation}
    \exp\left[{-i\sum_{j=1}^{m}H_{j}t}\right] = \prod_{j=1}^m \exp\left[{-i H_{j}t}\right] + \textit{O}(m^2 t^2),
    \label{trotter}  
\end{equation}
where $m$ is the number of time-steps. 

Noting that Trotter decomposition is currently widely used in the framework of Quantum Approximate Optimization Algorithm (QAOA), which uses both quantum and classical computer resources, see, e.g.,~\cite{farhi2014, LukinQAOA}. In practice, if the number of variational parameters is large enough, QAOA can solve the MaxCut problem with high enough accuracy.   

Ideally, a larger number of time-steps in the decomposition leads to more accurate results. But in our case, we have to keep in mind the depth of the circuit, as a larger circuit depth can lead to an accumulation of errors. In our Trotter decomposition, we keep the number of time-steps at $m = 6$ to have a smaller circuit depth without compromising the accuracy of the results. In general, as the dimension of the hypercube increases, the circuit depth increases by a factor of two gates. The corresponding quantum circuit obtained for the CTQW simulation is given in Fig.~\ref{fig4}(c). The circuit consists of $R_z$ and $\sqrt{X}$ gate. The $R_z$ is defined in Eq.~(\ref{R_z}) as,
\begin{equation}
     R_z(\theta) = \begin{bmatrix}
    \mathrm{e}^{-i\frac{\theta}{2}} &~~0\\\\
    0 &~~\mathrm{e}^{i\frac{\theta}{2}}
    \end{bmatrix},
    \label{R_z}
\end{equation}

The $\sqrt{X}$ gate is defined in Eq.~(\ref{sqrt(X)_gate}) as,
\begin{equation}
     \sqrt{X} = \begin{bmatrix}
    1+i &~~1-i\\\\
    1-i &~~1+i
    \end{bmatrix},
    \label{sqrt(X)_gate}
\end{equation}

In this mapping, the nodes of the line are encoded using one-hot encoding. In this encoding, the states are represented by bit strings, which consist of ``$1$'' at nodes where the particle can be found and zeros elsewhere. Therefore, the total Hilbert space of the CTQW gets reduced to these states. The described encoding helps in error-correcting all the other states those are not valid in the one-hot encoding, which we obtain while implementing the CTQW on IBM Q devices.

For $n=3$ hypercube, which is shown in Fig.~\ref{fig4}(a), we can compare this evolution of the CTQW mapped to a line with the CTQW implemented on individual qubits as shown in Fig.~\ref{fig2}(c). We observe that the walk's evolution on each time step is similar for both implementations if we correct the experimental errors occurring during the implementation. The quantum and classical simulation results are shown in Fig.~\ref{fig4}(d). Compared to the simulation results in Fig.~\ref{fig2}(c), the probabilities mismatch has two origins: errors because of the Trotterization procedure, and error because of the larger depth of the experimental quantum circuit. In both Fig.~\ref{fig2} and Fig.~\ref{fig4}, the transport between opposite hypercube vertices should be noticed. In the case of the implementation shown in Fig.~\ref{fig4}, also corresponds to transport in quantum spin networks~\cite{christandl2004perfect}.

\subsection{Simulating Arbitrary Graphs using Quantum Circuits}

 In this section, we discuss how an arbitrary graph can be mapped to a quantum circuit. We know that the adjacency matrix of a graph corresponds to the Hamiltonian which governs the state evolution of the quantum particle on the graph. Therefore, in order to simulate a graph on a quantum computer, we need to find an efficient way to decompose the unitary matrix which corresponds to the state evolution of the Hamiltonian into Pauli matrices. Hamiltonian simulation method finds an efficient decomposition of the unitary state evolution of a Hamiltonian into a product of Pauli terms which in turn, can be simulated in a quantum computer by using gates which are native to the architecture. We can use the Hamiltonian simulation technique to simulate a graph on a quantum computer by mapping the graph to a quantum circuit. 
 
 \begin{figure*}[ht!]
	\centering
	\includegraphics[width=1\linewidth]{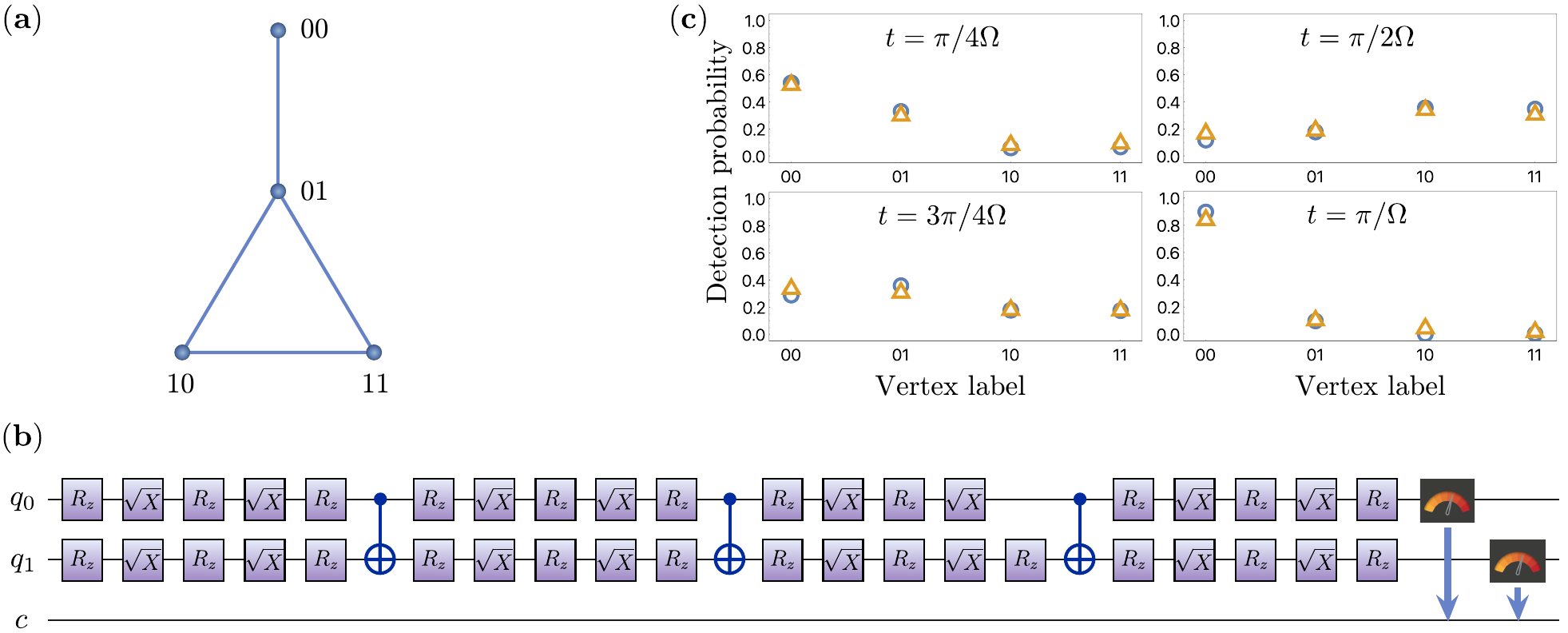}
	\caption{(a) A 'paw' graph with 4 nodes labelled in order is shown (b) Quantum circuit for a CTQW implemented for $t=\pi/2\Omega$ on the 'paw' graph. (c) Quantum and classical simulation results for implementing a CTQW 'paw' graph with $4$ vertices. Each plot shows a probability distribution of particles position on a graph at different times: $t=\pi/4\Omega$, $\pi/2\Omega$, $3\pi/4\Omega$, and $\pi/\Omega$. The blue circles and orange triangles correspond to execution on the simulator and the IBM Q quantum register, respectively. Vertex labels indicate the labelling of different vertices of the graph.}
	\label{fig5}
\end{figure*}

For example, we simulated a ``paw" graph with 4 nodes as shown in Fig.5. Also, 
shown is the quantum circuit obtained from Hamiltonian simulation using Qiskit implemented on the IBM quantum computer. The feasibility of implementing an arbitrary graph on a NISQ device can be evaluated by estimating the complexity of the Hamiltonian simulation method used. The query complexity of the Hamiltonian simulation directly depends on the size or complexity of the graph. We will discuss more about Hamiltonian simulation methods and their complexity in the next section.

\begin{table}[ht!]
\renewcommand{\arraystretch}{1.5}
\centering
\begin{tabularx}{0.48\textwidth} {
  | >{\centering\arraybackslash}X 
  | >{\centering\arraybackslash}X | }
\hline
 Algorithms & Complexity  \\ 
 \hline\hline
Product formulas \cite{berry2007efficient, childs2010simulating} & $\textit{O}(d^3t(dt/\epsilon)^{1/2k})$  \\ [2ex]
Phase estimation on Quantum Walks (QW) \cite{childs2010relationship, berry2009black} & $\textit{O}(dt/\sqrt{\epsilon})$ \\ [4ex]
Fractional queries \cite{berry2017exponential} or Truncated Taylor series \cite{berry2015simulating} & $\textit{O}(d^2t\frac{\log(d^2t/\epsilon)}{\log\log(d^2t/\epsilon)})$ \\ [4ex]
Linear combination of QW \cite{berry2015hamiltonian} & $\textit{O}(dt\frac{\log(dt/\epsilon)}{\log\log(dt/\epsilon)})$ \\[4ex]
Quantum Signal Processing \cite{low2017optimal} & $\textit{O}(dt + \log(1/\epsilon))$ \\[4ex]
Qubitization/ QSP \cite{low2019hamiltonian}& $\textit{O}(d^2t + \log(1/\epsilon))$ \\ [4ex]
 \hline
\end{tabularx}

\caption{Algorithms for Hamiltonian simulation along with their query complexity where $k$ is the sparcity, $t$ is the time parameter and $\epsilon$ is the error parameter. Here QW refers to Quantum walks and QSP refers to Quantum Signal Processing respectively.}
\end{table}

\subsection{Hamiltonian simulation} 
Hamiltonian simulation methods in quantum information science addresses the problem of efficient simulation of quantum systems. The goal of the algorithm is to find an approximation to a unitary matrix U such that, $||U - \mathrm{e}^{-i\Omega t A}|| \leq \epsilon$ where $\epsilon$ is the maximum simulation error and $||.||$ is the spectral norm. There are different algorithms or techniques with varying complexity used for simulating the unitary state evolution of a Hamiltonian in a quantum computer. They can be mainly divided into divide and conquer algorithms and quantum walk algorithms. We can use Hamiltonian simulation techniques to simulate an arbitrary graph on a quantum computer. The complexity of the algorithm determines how efficiently we can simulate graphs on a quantum computer. Therefore, larger complex graphs are more challenging to be simulated on quantum computers. Thus creating the need to improve the performance of current Hamiltonian simulation algorithms more important. Some of the most prominent algorithms for Hamiltonian simulation along with their query complexities are given in Table 1.

\section{Conclusion}
We demonstrate the feasibility of NISQ device for implementation of CTQW. Quantum walks are implemented utilizing classical and quantum simulation, where quantum simulations are performed on currently available quantum computers of IBM Q. All quantum walks are implemented with the number of qubits that scale logarithmically with the graph size.

The presented results consist of two parts. First, a mapping between circuits and graphs is shown. A method to obtain the CTQW on a graph corresponding to a quantum circuit is discussed. The quantum circuit is simulated on the state-of-the-art IBM quantum computer to obtain the graphs. Hypercube graphs and arbitrary graphs are obtained. In the second part, a method to map graphs to the circuit is studied. Hypercubes and arbitrary graphs are simulated on a quantum computer using quantum circuits. Both classical and quantum simulation results are obtained and compared. The technique of Hamiltonian simulation is discussed and various algorithms for Hamiltonian simulation are listed with their complexities.

With our work, we hence established an analogy between quantum circuits and graphs that allowed us to tackle computationally challenging simulation problems. This result paves the way towards the practical realization of quantum advantage in quantum walk simulation for algorithm development.

\medskip
Author contribution statement:

Anandu Kalleri Madhu and Alexey Melnikov: Conceived and designed the experiments

Anandu Kalleri Madhu and Alexander Alodjants: Contributed analysis tools and data

Anandu Kalleri Madhu: Performed the experiments

Anandu Kalleri Madhu, Alexey Melnikov, Leonid Fedichkin, Alexander Alodjants, Ray-Kuang Lee: Analyzed and interpreted the data; Wrote the paper

\medskip
Funding statement:

Leonid Fedichkin was supported by Ministry of Science and Higher Education of Russia for Valiev Institute of Physics and Technology of RAS [Program No. FFNN-2022-0016].

Alexander Alodjants was supported by Ministry of Science and Higher Education of the Russian Federation and South Ural State University [No. 075-15-2022-1116].

Ray-Kuang Lee was supported by Ministry of Science and Technology of Taiwan [No. 109-2112- M-007-019-MY3, 109-2627-M-008-001, 110-2123-M-007-002].

\medskip
Data availability statement:

Data associated with this study has been deposited at https://github.com/q-ml/quantum-walks-quantum-devices

\medskip
Declaration of interest’s statement:

The authors declare no competing interests.

\bibliography{qw_ibm}

\section{Supplemental information}

In this section we provide two tables of parameters, Table.~\ref{tab:fig4_gate labels} and Table.~\ref{tab:fig5_gate labels} for the quantum circuit used in Fig.~\ref{fig4}(c) and Fig.~\ref{fig5}(b) respectively.

All $4$-qubit circuits were executed on the ``ibmq\_bogota'' IBM quantum processor, and all $20$-qubit circuits were executed on the ``ibmq\_paris'' IBM quantum processor.

All the codes and data related to this work is provided in this public GitHub repository: https://github.com/q-ml/quantum-walks-quantum-devices

\begin{table*}[htp]
\centering
\footnotesize\setlength{\tabcolsep}{2.5pt}
\begin{tabular}{l@{\hspace{4pt}} *{23}{c}}
\toprule
\bfseries Parameters & \multicolumn{20}{c}{\bfseries $\bm R_z$ Gates} \\
& 1 & 2 & 3 & 4 & 5 & 6 & 7 & 8 & 9 & 10 & 11 & 12 & 13 & 14 & 15 & 16 & 17 & 18 & 19 & 20 & 21 & 22 & 23 \\
\bfseries $\theta$ & -1.57 & -0.64 & 1.57 & -0.64 & 0.79 & 4.18 & -2.37 & 4.18 & 8.82 & 8.82 & -1.86 & -1.86 & 1.57 & 1.57 & 4.76 & 4.76 & 11.3 & 11.3 & 0.79 & -1.05 & 0.79 & -1.05 & -3.14 \\
\hline
\bfseries  & 24 & 25 & 26 & 27 & 28 & 29 & 30 & 31 & 32 & 33 & 34 & 35 & 36 & 37 & 38 & 39 & 40 & 41 & 42 \\

\bfseries $\theta$  & -1.76 & -3.14 & -1.76 & 2.09 & 2.09 & -1.86 & -1.86 & 1.57 & 1.57 & 4.76 & 4.76 & 11.3 & 11.3 & -3.14 & -1.57 & -3.14 & -1.57 & 0.76 & -2.38\\
\end{tabular}
\caption{The $R_z$ gate parameters used in the circuit in Fig.~\ref{fig4}(c) are given in the order from top to bottom, left to right.}\label{tab:fig4_gate labels}
\end{table*}

\begin{table*}[htp]
\centering
\footnotesize\setlength{\tabcolsep}{2.5pt}
\begin{tabular}{l@{\hspace{4pt}} *{23}{c}}
\toprule
\bfseries Parameters & \multicolumn{20}{c}{\bfseries $\bm R_z$ Gates} \\
& 1 & 2 & 3 & 4 & 5 & 6 & 7 & 8 & 9 & 10 & 11 & 12 & 13 & 14 & 15 & 16 & 17 & 18 & 19 & 20 & 21 & 22 & 23 \\
\bfseries $\theta$ & -0.12 & -0.49 & 4.59 & 4.32 & 11.74 & 11.15 & -1.57 & 1.5 & 3.46 & 5.52 & 7.85 & 12.51 & 1.57 & -2.56 & 2.85 & 4.2 & 10.07 & 2.32 & 2.02 & 4.59 & 5.32 & 12.44 & 7.75  \\

\end{tabular}
\caption{The $R_z$ gate parameters used in the circuit in Fig.~\ref{fig5}(b) are given in the order from top to bottom, left to right.}\label{tab:fig5_gate labels}
\end{table*}
\end{document}